\documentclass[twocolumn,aps,prl,showpacs,preprintnumbers,amsmath,amssymb, superscriptaddress]{revtex4-1}
\usepackage{bm}
\usepackage{graphicx}
\usepackage{xcolor,xspace}
\usepackage{ulem}
\usepackage{float}

\begin{document}

\title{Spontaneous gap opening and potential excitonic states in an ideal Dirac semimetal Ta$_2$Pd$_3$Te$_5$ }

\author{Peng~Zhang}\thanks{These authors contributed equally to this work.}
\affiliation{School of Physics and National Laboratory of Solid State Microstructures, Nanjing University, Nanjing, China}
\affiliation{Collaborative Innovation Center for Advanced Microstructures, Nanjing, China}
\email{zhangpeng@nju.edu.cn}

\author{Yuyang~Dong}\thanks{These authors contributed equally to this work.}
\affiliation{Institute for Solid State Physics, the University of Tokyo, Kashiwa, Chiba 277-8581, Japan}

\author{Dayu~Yan}\thanks{These authors contributed equally to this work.}
\affiliation{Beijing National Laboratory for Condensed Matter Physics, Institute of Physics, Chinese Academy of Sciences, Beijing, China}
\affiliation{University of Chinese Academy of Sciences, Beijing, China}

\author{Bei~Jiang}\thanks{These authors contributed equally to this work.}
\affiliation{Beijing National Laboratory for Condensed Matter Physics, Institute of Physics, Chinese Academy of Sciences, Beijing, China}
\affiliation{University of Chinese Academy of Sciences, Beijing, China}

\author{Tao~Yang}
\affiliation{School of Physics and National Laboratory of Solid State Microstructures, Nanjing University, Nanjing, China}

\author{Jun~Li}
\affiliation{Beijing National Laboratory for Condensed Matter Physics, Institute of Physics, Chinese Academy of Sciences, Beijing, China}

\author{Zhaopeng~Guo}
\affiliation{Beijing National Laboratory for Condensed Matter Physics, Institute of Physics, Chinese Academy of Sciences, Beijing, China}
\affiliation{University of Chinese Academy of Sciences, Beijing, China}

\author{Yong~Huang}
\affiliation{School of Physics and National Laboratory of Solid State Microstructures, Nanjing University, Nanjing, China}

\author{Bo~Hao}
\affiliation{College of Engineering and Applied Sciences, Nanjing University, Nanjing, China}

\author{Qing~Li}
\affiliation{School of Physics and National Laboratory of Solid State Microstructures, Nanjing University, Nanjing, China}

\author{Yupeng~Li}
\affiliation{Beijing National Laboratory for Condensed Matter Physics, Institute of Physics, Chinese Academy of Sciences, Beijing, China}

\author{Kifu~Kurokawa}
\affiliation{Institute for Solid State Physics, the University of Tokyo, Kashiwa, Chiba 277-8581, Japan}

\author{Rui~Wang}
\affiliation{School of Physics and National Laboratory of Solid State Microstructures, Nanjing University, Nanjing, China}
\affiliation{Collaborative Innovation Center for Advanced Microstructures, Nanjing, China}

\author{Yuefeng~Nie}
\affiliation{College of Engineering and Applied Sciences, Nanjing University, Nanjing, China}

\author{Makoto~Hashimoto}
\affiliation{Stanford Synchrotron Radiation Lightsource, SLAC National Accelerator Laboratory, Menlo Park, CA, USA}

\author{Donghui~Lu}
\affiliation{Stanford Synchrotron Radiation Lightsource, SLAC National Accelerator Laboratory, Menlo Park, CA, USA}

\author{Wen-He Jiao}
\affiliation{Key Laboratory of Quantum Precision Measurement of Zhejiang Province, Department of Applied Physics, Zhejiang University of Technology, Hangzhou, China} 

\author{Jie~Shen}
\affiliation{Beijing National Laboratory for Condensed Matter Physics, Institute of Physics, Chinese Academy of Sciences, Beijing, China}

\author{Tian~Qian}
\affiliation{Beijing National Laboratory for Condensed Matter Physics, Institute of Physics, Chinese Academy of Sciences, Beijing, China}
\affiliation{University of Chinese Academy of Sciences, Beijing, China}

\author{Zhijun~Wang}
\affiliation{Beijing National Laboratory for Condensed Matter Physics, Institute of Physics, Chinese Academy of Sciences, Beijing, China}
\affiliation{University of Chinese Academy of Sciences, Beijing, China}

\author{Youguo~Shi}
\affiliation{Beijing National Laboratory for Condensed Matter Physics, Institute of Physics, Chinese Academy of Sciences, Beijing, China}
\affiliation{University of Chinese Academy of Sciences, Beijing, China}

\author{Takeshi~Kondo}
\affiliation{Institute for Solid State Physics, the University of Tokyo, Kashiwa, Chiba 277-8581, Japan}
\affiliation{Trans-scale Quantum Science Institute, The University of Tokyo, Tokyo 113-0033, Japan}
\email{kondo1215@issp.u-tokyo.ac.jp}

\date{\today}

\begin{abstract}

\textbf{
The opening of an energy gap in the electronic structure generally indicates the presence of interactions. In materials with low carrier density and short screening length, long-range Coulomb interaction favors the spontaneous formation of electron-hole pairs, so-called excitons, opening an excitonic gap at the Fermi level. Excitonic materials host unique phenomenons associated with pair excitations. 
However, there is still no generally recognized single-crystal material with excitonic order, which is, therefore, awaited in condensed matter physics. Here, we show that excitonic states may exist in the quasi-one-dimensional material Ta$_2$Pd$_3$Te$_5$, which has an almost ideal Dirac-like band structure, with Dirac point located exactly at Fermi level. We find that an energy gap appears at 350 K, and it grows with decreasing temperature. The spontaneous gap opening is absent in a similar material Ta$_2$Ni$_3$Te$_5$. Intriguingly, the gap is destroyed by the potassium deposition on the crystal, likely due to extra-doped carriers. Furthermore, we observe a pair of in-gap flat bands, which is an analog of the impurity states in a superconducting gap. All these observations can be properly explained by an excitonic order, providing Ta$_2$Pd$_3$Te$_5$ as a new and promising candidate realizing excitonic states.
}

\end{abstract}

\maketitle

When the dimensionality is reduced, the screening of the Coulomb interaction becomes weak, leading to strong electron correlations in favor of various novel states. One such example is an excitonic state, which hosts electron-hole pairs bounded by Coulomb attraction, namely the excitons. The excitonic order builds up spontaneously when the binding energy of excitons is larger than the original band gap~\cite{KohnPR1967}. 
Similar to superconductors, excitonic insulators open an energy gap corresponding to the binding energy of excitons and have a pair-breaking effect by impurity scattering~\cite{ZittartzPR1967}. Contrasting to a large variety of superconducting materials, however, excitonic insulator candidates are very rare. 

Excitonic states have been so far reported for some of fabricated bilayer structures~\cite{EisensteinNature2004,DuNC2017,DeanNP2017,ShanNature2021,DeanScience2022,WangNature2023} and for monolayer tellurides~\cite{WuNP2022,CobdenNP2022,ChenNC2023,ShenNC2023}.
There are also reports for single crystals but only of two compounds under normal conditions: \textit{1T}-TiSe$_2$ and Ta$_2$NiSe$_5$~\cite{CercellierPRL2007,MonneyPRB2009,BeckPRL2011,AbbamonteScience2017,TakagiPRL2009,OhtaPRL2018,RaoSA2021,YeomNP2021,KimNC2021}. These two single-crystal candidates, however, entangle with other phase transitions, making the verification of the possible excitonic states rather complex. In \textit{1T}-TiSe$_2$, the valence and conduction bands are located at different momentum spaces with an indirect gap. The excitonic pairing with a finite wavevector naturally coexists with charge-density-wave (CDW), leading to intrinsic difficulties in distinguishing excitonic order from CDW. 
In Ta$_2$NiSe$_5$, both the valence band and conduction band locate at the Brillouin zone center with a direct gap. In such a case, there is no coexisting CDW. However, an orthorhombic-to-monoclinic structure transition appears at the proposed excitonic transition temperature. Whether the change of energy gap is induced by excitonic order or structure transition, is still under debate~\cite{KingPRR2020,GedikarXiv2020,HearXiv2022}. Because of these circumstances, one has been awaiting for single crystals which host only excitonic order and do not entangle with other phase transitions. 
In particular, theoretical calculations~\cite{WangarXiv2024} predict Ta$_2$Pd$_3$Te$_5$ monolayer to be an excitonic insulator with rather simple band structures, inspiring experiments to explore the possible excitonic states in the single crystals of this compound.  

\begin{figure*}[!t]
\begin{center}
\includegraphics[width=0.95\textwidth]{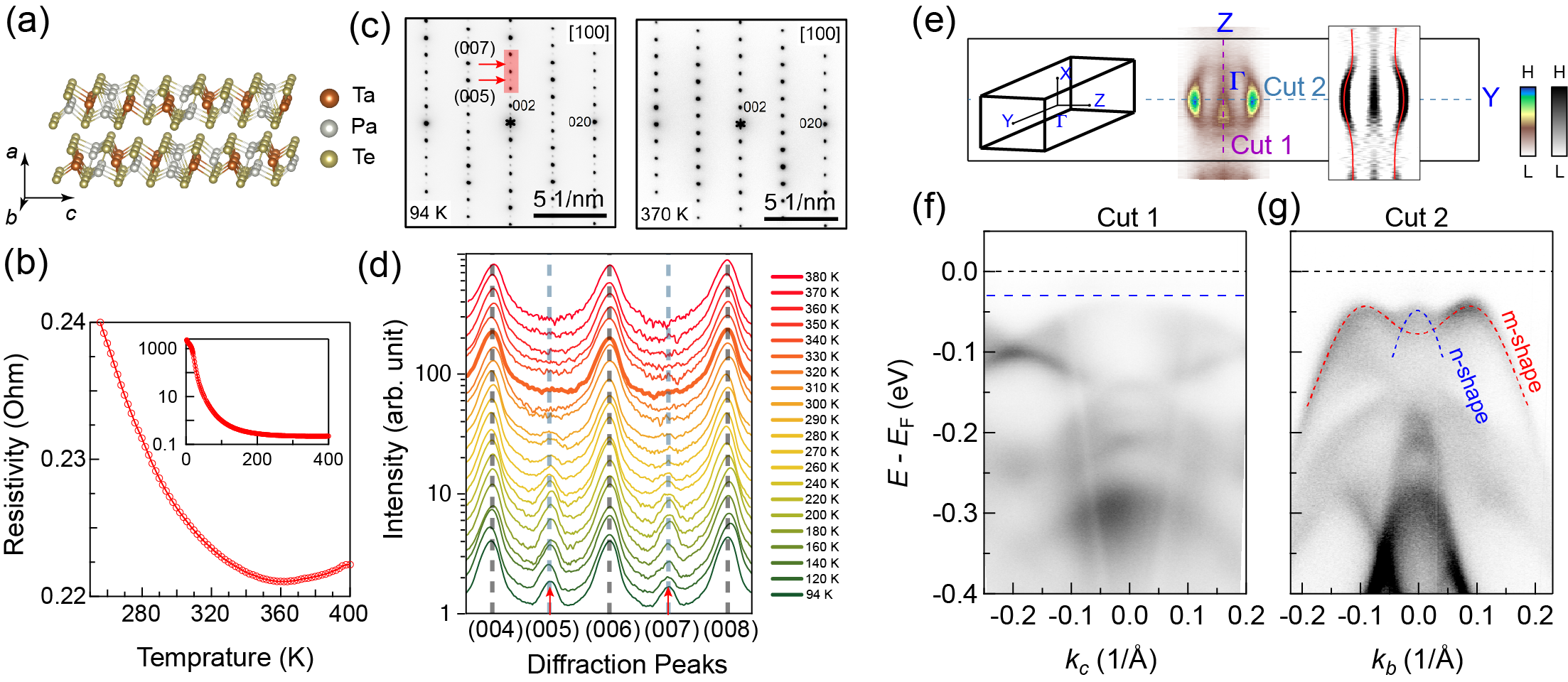}
\end{center}
 \caption{\label{band}  \textbf{The crystal structure and electronic properties of Ta$_2$Pd$_3$Te$_5$.} (a) Crystal structure of Ta$_2$Pd$_3$Te$_5$. The Ta-Te chains are along the $b$ axis. (b) Resistivity curve zooming from 250 K to 400 K. The inset shows the overall behavior from 2K - 400 K. (c) Electron diffraction measurement of the [100] zone axis. The data are taken at 94 K and 370 K, respectively. (d) Temperature evolution of the (00l) peaks. The vertical intensity axis is in a log scale. See Supplementary Figure 1 for the same plot in a linear scale. (e)  Intensity mapping in $k$-space around the $\Gamma$-Z line measured by a 7-eV laser ARPES. The data is taken at -30 meV [dashed blue line in (f)] and integrated over an energy window of $\pm$5 meV. The left and right insets show the bulk Brillouin zone and the curvature plot of the data in greyscale, respectively.  (f,g) Band structures measured at 10 K by ARPES, perpendicular to the chain [Cut 1 in (e)] with $s$-polarized light and along the chain [Cut 2 in (e)] with $p$-polarized light, respectively.}
\end{figure*}

An ideal Dirac semimetal, in which the Dirac point is located exactly at the Fermi level, has very low carrier density (almost zero), and the electron and hole excitations are almost symmetric. In such a unique system, it is expected that the electrons and holes are easily bounded by Coulomb interaction, and an excitonic order naturally occurs. 
In this work, we find such an ideal Dirac semimetal state in single-crystal material Ta$_2$Pd$_3$Te$_5$, with the following features supporting the existence of excitonic order: (i) There is an energy gap opening at 350 K, and it is associated with a tiny lattice distortion unveiled by electron diffraction. (ii) The spontaneous gap opening is absent in a similar material Ta$_2$Ni$_3$Te$_5$. (iii) The energy gap is destroyed by carrier doping, according to our experiments of the potassium deposition on the crystal surface.  (iv) The in-gap spectra show double flat dispersions in the energy gap, most likely from impurity states, similar to those emerging in superconductors due to the breaking of Cooper pairs~\cite{SchriefferPRB1997,BalatskyRMP2006}. These observations put a strong limitation on the origin of the energy gap and are, most importantly,  explained reasonably by an excitonic order. 

Ta$_2$Pd$_3$Te$_5$ single crystals were synthesized by the self-flux method. Details of the synthesis process can be found in Ref~\citenum{WangPRB2021}. The high-resolution angle-resolved photoemission spectroscopy (ARPES) measurements were carried out with a 6.994-eV laser and a VG-Scienta R4000WAL electron analyzer at the Institute for Solid State Physics, the University of Tokyo. The energy resolution for the experiments was set to $\sim$ 3 meV. The synchrotron ARPES measurements were performed at beamline 5-2 of the Stanford Synchrotron Radiation Lightsource with a ScientaOmicron DA30L analyzer. The overall energy resolution was 10$\sim$20 meV. 
In both ARPES experiments, the spot size of incident light was smaller than $\sim$50 $\mu$m.  
The samples were cleaved \textit{in-situ}, and kept in an ultra-high vacuum below $5 \times 10^{-11}$ Torr during the ARPES measurements. 

Ta$_2$Pd$_3$Te$_5$ has a quasi-1D crystal structure~\cite{WangPRB2021,TakenakaJPSJ2021,FengPRB2021,JiaoAPS2022}, with Ta-Te or Pd-Te chains along the $b$ axis [Fig. 1(a)]. The resistivity measurement [Fig.1(b)] shows a typical semiconducting behavior at low temperatures. It shows an upturn around 350-360 K and becomes metallic-like above 350-360 K, indicating some changes in the electronic structure. Indeed, electron diffraction experiments unveil some structure distortions. As shown in Fig. 1(c), at 370 K the electron diffraction data taken along the [001] zone axis consists of (hkl) peaks with $k+l = 2n$ (n is an integer), consistent with that of space group No. 62, while at 94 K the peaks with $k+l=2n+1$ appear. From the temperature evolution in Fig.1(d), we find the distortion happens around 330 K [thick line in Fig. 1(d)], roughly consistent with the transition temperature in resistivity measurement. 
We note the emerged $k+l=2n+1$ peaks at 94 K are very weak. For example, the (005) peak intensity is just 2.7\% of (004) peak intensity in Fig.1(c). This indicates a very small lattice distortion. 
Indeed, X-ray Diffraction (XRD) along (h00) and specific heat measurement did not resolve such small structure distortion (Supplementary Figure 2), probably because they are not as efficient as electron diffraction. More sophisticated measurements may be able to resolve this tiny lattice distortion from XRD and specific heat. 

In Fig. 1(e), we plot the ARPES intensity map measured at 10K along the $\Gamma$-Y-Z sheet slightly below the Fermi level ($E_\mathrm{F}$) at -30 meV. The quasi-1D character of the electronic structure is evident from parallel segments with only a little warping. 
This feature is more clearly demonstrated in the right inset panel by taking the curvature of the ARPES intensities to show up the location of energy states~\cite{ZhangRSI2011}.  Figures 1(f) and 1(g) display the band structures along the high symmetry lines $\Gamma$-Z and $\Gamma$-Y, described as Cut 1 and Cut 2 in Fig. 1(e), respectively.  The energy dispersion is much stronger along Cut 2 than along Cut 1 due to the quasi-1D structure. 
Along Cut 2, we observe three humps just below $E_\mathrm{F}$, consisting of an m-shape band and an n-shape band in the middle, as illustrated in Fig. 1(g). 

\begin{figure*}[!htbp]
\begin{center}
\includegraphics[width=0.9\textwidth]{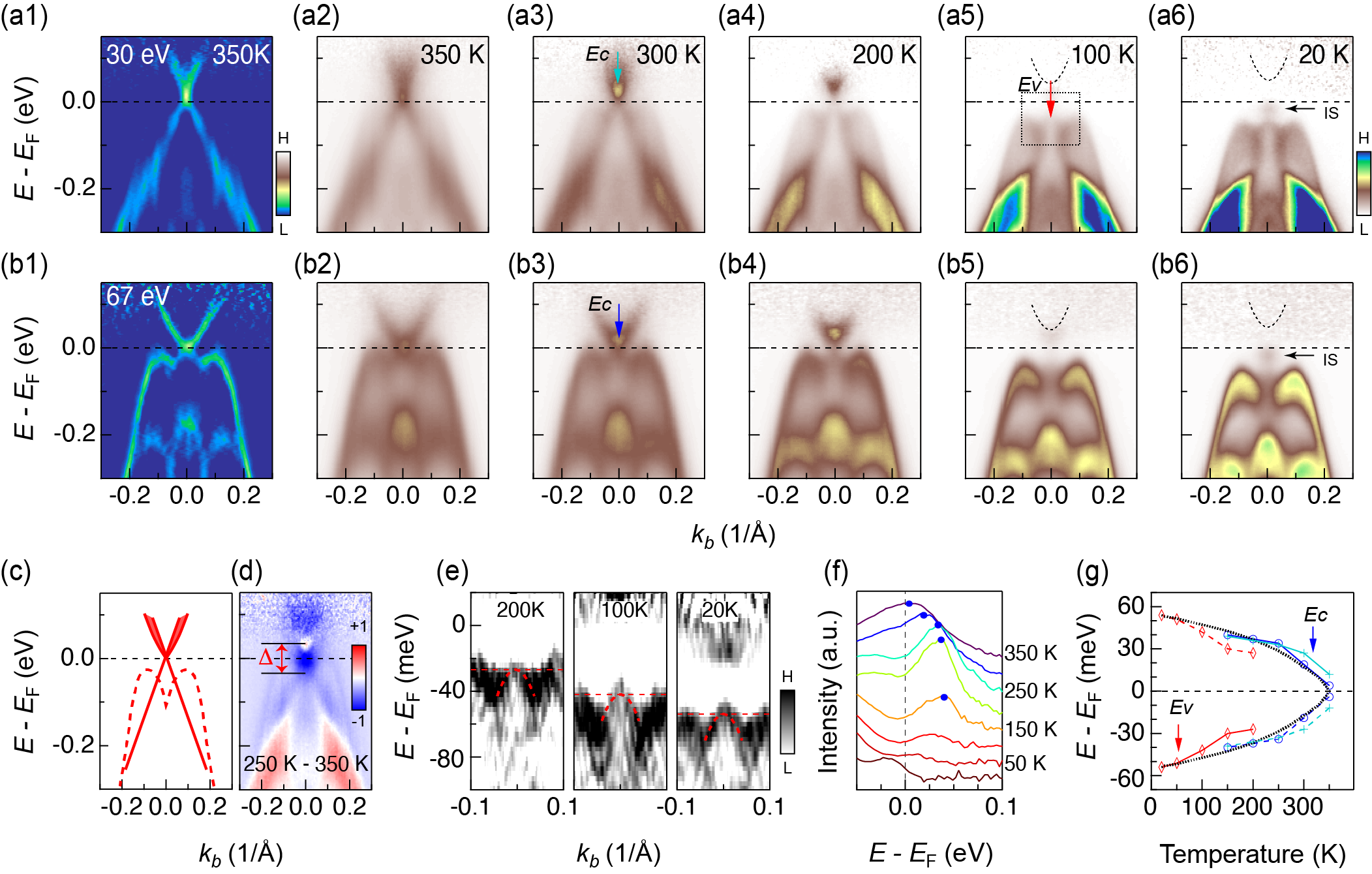}
\end{center}
 \caption{\label{spin} \textbf{Temperature evolution of the band structure and energy gap in Ta$_2$Pd$_3$Te$_5$.} (a1) The curvature plot of (a2). 
 (a2-a6) The electronic structure along the chain direction at $k_c=0$, measured by ARPES from 350 K to 20 K with 30-eV photons. 
 (b1) The curvature plot of (b2). (b2-b6) The same data as (a2-a6), but measured with 67-eV photons. The black dashed lines in (a5-a6) and (b5-b6) sketch the conduction band above $E_\mathrm{F}$. The black arrows in (a6) and (b6) mark the in-gap impurity states (IS), which do not cross Fermi level [See their high resolution EDCs in Fig.5(d)]. The details of impurity states are discussed in Fig.5.
 All these data are divided by the Fermi function at measured temperatures to remove the cut-off effect above $E_\mathrm{F}$. 
 (c) Sketch of band dispersions extracted from 30-eV and 67-eV data. The cone-shaped bands crossing $E_\mathrm{F}$ and the m-shape band not crossing $E_\mathrm{F}$ are represented by solid and dashed lines, respectively. The conduction bands from 30-eV and 67-eV data have slightly different dispersions. 
 (d) Intensity difference between 250 K and 350 K data measured with 30-eV photons. The blue area around $E_\mathrm{F}$ indicates the opening of the energy gap $\Delta$. 
 (e)  EDC curvature plot for the 30-eV data zooming the valence band top [dashed box in (a5)] for three different temperatures (200K, 100K, and 20K).
The valence band top [red arrow in (a5)] is extracted from the curvature-enhanced dispersion. 
 (f) EDCs of the conduction band at $k_b=k_c=0$, extracted from 67-eV data for different temperatures. The blue dots indicate the conduction band bottom [blue arrow in (b3)]. 
 (g) Temperature dependence of valence band top ($E_v$) and conduction band bottom ($E_c$), demonstrating the gap evolution. 
 The $E_v$ (solid red line) is extracted from (e), and $E_c$ (solid blue line) is extracted from (f). The $E_c$ curve in cyan color is extracted from EDCs in 30-eV data in (a).
As both $E_c$ and $E_v$ values are acquired at 150 K and 200 K and they are roughly equal, it is reasonable to assume that the band gap opens symmetrically to $E_\mathrm{F}$. Then we get the full temperature evolution of conduction and valence bands by mirroring the blue, cyan and red lines. The entire evolution of the gap with temperature is illustrated by the dashed black lines. }
 \end{figure*}

We investigate the evolution of the band structure over a wide range of temperatures from 350 K to 20 K, with 30-eV and 67-eV photons, as displayed in Fig. 2(a-b). Here the data are divided by the Fermi function at measurement temperatures to remove the cut-off effect above $E_\mathrm{F}$. At 350 K [Figs. 2(a2) and 2(b2)], Dirac-type bands are observed. It is further clarified in Fig. 2(a1) and Fig. 2(b1) by taking the curvature of those~\cite{ZhangRSI2011}. The Dirac point is located almost exactly at the Fermi level, indicating an ideal Dirac semimetal state. The band dispersions at different photon energies are extracted and illustrated in Fig. 2(c). Except for one m-shape valence band and one cone-shape (n-shape at low temperatures) valence band, the conduction bands from 30-eV and 67-eV data show slightly different dispersions. These two dispersions may come either from the same band at different $k_a$ (out of plane direction, or $k_z$ in general expression for ARPES) or different bands. We cannot distinguish these two scenarios due to the limited momentum resolution along $k_a$ in ARPES. The first-principles calculations (Supplementary Figure 3) show only one conduction band with a small $k_a$ dispersion around $E_\mathrm{F}$; this suggests that conduction bands of the 30-eV and 67-eV data are likely originated from the same band and exhibits slightly different dispersions due to different $k_a$. 

\begin{figure*}[!htbp]
\begin{center}
\includegraphics[width=0.75\textwidth]{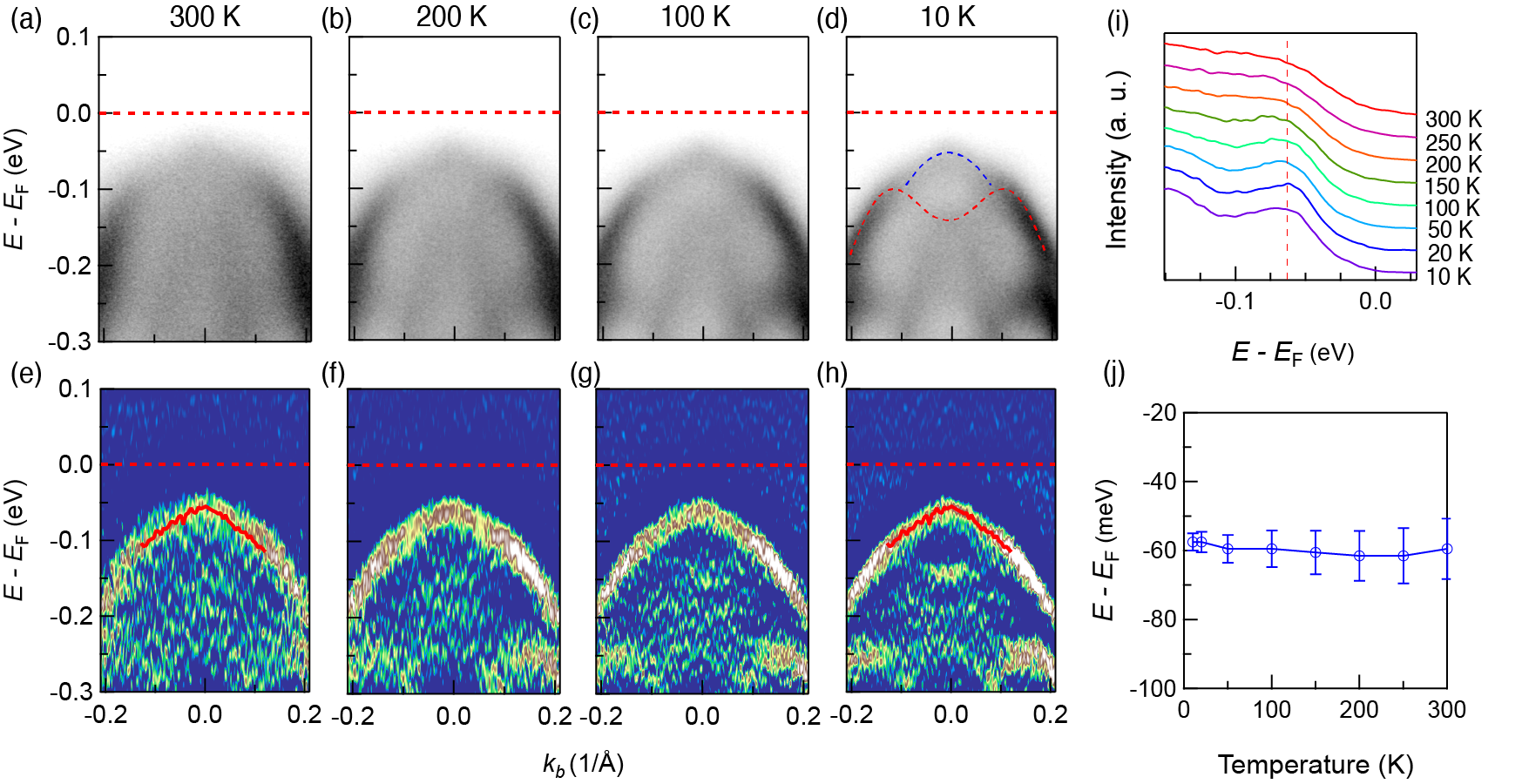}
\end{center}
 \caption{\label{spin} \textbf{Temperature evolution of the band structure in Ta$_2$Ni$_3$Te$_5$.} (a - d) The electronic structure along the chain direction at $k_c = 0$, measured from 300 K to 10 K with 7-eV photons. (e-h) The curvature plot of (a - d). The red lines in (e) and (h) are the same curve, extracted from EDC peaks in (h). (i) Evolution of EDCs with temperature at $k_b = 0$. (j) Shift of valence band top with temperature. Blue dots are extracted from EDC peaks in (e - h). }
\end{figure*}

Since the m-shape valence band does not cross $E_\mathrm{F}$, we focus on the cone-shape (n-shape at low temperatures) valence band and the conduction band crossing $E_\mathrm{F}$.  With decreasing temperature, an energy gap gradually develops between the conduction and valence bands in both Fig. 2(a2-a6) and Fig. 2(b2-b6). The gap-opening feature is further demonstrated in Fig. 2(d) by plotting  
the intensity difference between the data at 250 K and 350 K for 30 eV. The reduced intensity (blue color) due to a gap opening ($\Delta$) is seen around $E_\mathrm{F}$. We note the intensities just below $E_\mathrm{F}$ in Fig.2(a6) and Fig.2(b6) are from impurity states (IS), which are discussed in detail in Fig.5.

We examine the gap behavior in more detail by tracing the temperature evolution of the valence band top ($E_v$) and the conduction band bottom ($E_c$). We trace $E_v$ from 30-eV data and  $E_c$ from 30-eV and 67-eV data [see cyan, red and blue arrows in Fig. 2(a3), Fig. 2(a5) and Fig. 2(b3), respectively]. The $E_v$ is obtained from the curvature-enhanced intensity plot in Fig. 2(e), which correspond to the dash box region in Fig. 2(a5). We see the shifting down of $E_v$ from -27 meV at 200 K to -54 meV at 20 K. Above 200 K, however, the valence band top has some intensity overlap with the conduction band bottom, so the $E_v$ cannot be accurately determined. The temperature evolution of $E_v$ from 30-eV data is plotted with a solid red line in Fig. 2(g) for the range below 200K. We determine the $E_c$ at 67 eV directly from the peak positions of the energy distribution curves (EDCs), as shown in Fig. 2(f) with blue dots marking the peaks. The evolution of $E_c$ from 67-eV data with temperature is plotted with a solid blue line in Fig. 2(g). The cyan $E_c$ curve is acquired in a similar way, but from 30-eV data.  Note that the plot is only for the range above 150 K, since the band bottom cannot be reliably extracted below 150 K due to the cut-off effect of the Fermi function.

Both $E_c$ and $E_v$ values are acquired at 150 K and 200 K and they are roughly equal. Thus, it would be reasonable to assume that the band gap opens symmetrically to $E_\mathrm{F}$, $E_c \approx E_v$. With this assumption, we can compensate the missing values of $E_c$ and $E_v$ 
at low and high temperatures, respectively, by mirroring the solid red and blue lines according to $E_\mathrm{F}$, as shown in Fig. 2(g). The difference between the mirrored and original curves in the overlapped temperature range (150 K and 200 K) is small enough to justify the assumption of symmetric gap opening to $E_\mathrm{F}$. The dashed black lines sketch the full evolution of the conduction and valence bands, and thus the gap with temperature. From this result, we conclude that the gap has a value of $\sim$ 100 meV at 20 K, and closes at about 350 K.

As a comparison, we further measured the band structure of Ta$_2$Ni$_3$Te$_5$, which has the same crystal structure as Ta$_2$Pd$_3$Te$_5$. At low temperature 10 K [Fig.3(d)], the band structure of Ta$_2$Ni$_3$Te$_5$ also shows a m-shape band [Red line in Fig.3(d)] and a parabolic band [Blue line in Fig.3(d)] near $E_F$, similar to the one in Fig.1(g).
In sharp contrast to the spontaneous gap opening in Ta$_2$Pd$_3$Te$_5$, we notice the valence band in Ta$_2$Ni$_3$Te$_5$ does not shift with temperature at all, as shown in Fig.3(a-d) and their curvature plot Fig.3(e-h). The valence band dispersion at 10 K are extracted from EDC peaks in Fig.3(h) and plotted in both Fig.3(e) and Fig.3(h). We find the dispersion from 10 K overlaps with the 300 K data very well, indicating the valence band has no change from 300 K to 10 K. The EDCs at $k_b = 0$ at different temperatures are plotted in Fig.3(i). As guided by the red dashed line in Fig.3(i), the EDC peaks show no change over a temperature range of 10 - 300 K. We also extract the EDC peaks in Fig.3(e - h), and plot the peak position in blue line in Fig.3(j). The curve is almost flat, further demonstrating the absence of energy gap change. 
Ta$_2$Ni$_3$Te$_5$ is a semiconductor, while Ta$_2$Pd$_3$Te$_5$ is a semimetal at 350 K. At the same time, the spontaneous gap opening is absent in Ta$_2$Ni$_3$Te$_5$ but exists in Ta$_2$Pd$_3$Te$_5$. Since the two materials have exactly the same crystal structure, we conclude that the spontaneous gap opening with temperature is originated from the semimetal nature of Ta$_2$Pd$_3$Te$_5$, and it is not a simple conduction band and valence band shift with temperature as in some semiconductors~\cite{PasslerPRB2002,CohenPRL2010,YangJPCM2020}.

\begin{figure*}[!htb]
\begin{center}
\includegraphics[width=.85\textwidth]{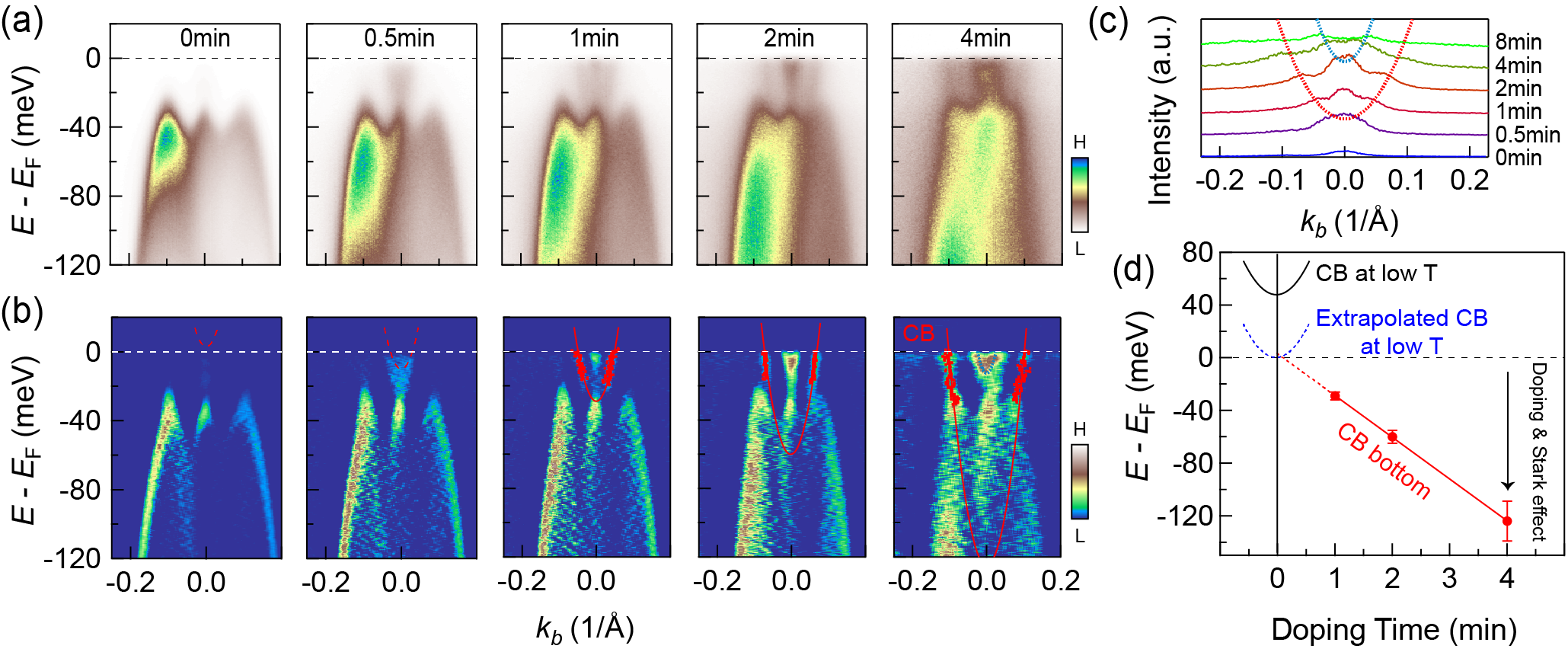}
\end{center}
 \caption{\label{theory} \textbf{Effect of potassium deposition on the band structure of Ta$_2$Pd$_3$Te$_5$. }  (a) Evolution of band structure along the chain direction at $k_c=0$ with potassium deposition, taken at 10 K. (b) Curvature intensity plot of (a). The energy states (small red dots) near $E_\mathrm{F}$ in 1-, 2- and 4-minute data are extracted from the peak positions of MDCs. The red lines are parabolic fits to the red dots. (c) Variations of MDCs at $E_\mathrm{F}$ with the time of potassium deposition. Dashed lines illustrate the change of the Fermi vector $k_\mathrm{F}$ of the conduction bands. (d) Evolution of the conduction band (CB) with potassium deposition. The red dots represent the band bottom of the fitted bands in (b). The solid red line is a linear fit to the red dots, and the dashed red line is the extrapolation of it to 0 minutes. }
\end{figure*}

To distinguish the origin of the spontaneous gap opening in Ta$_2$Pd$_3$Te$_5$, we deposit potassium on the sample surface to dope extra carriers and investigate how the gap evolves. Figures 4(a) and 4(b) show the ARPES dispersions and the curvature intensities of those, respectively, measured at 10 K after deposition for different times from 0 up to 4 minutes (from left to right panels). 
The most pronounced variation is seen in the conduction band; its bottom shifts down below $E_\mathrm{F}$ after 0.5-minute deposition. 
With further deposition, another conduction band at higher energies appears below $E_\mathrm{F}$. 
The appearance of these conduction bands can also be identified in the momentum distribution curves (MDCs) at $E_\mathrm{F}$ [Fig. 4(c)]. From the MDC maximums, we determine the conduction band near $E_\mathrm{F}$ for 1-, 2- and 4-minute doping, and display those by red dots in Fig. 4(b). To analyze the conduction band shift quantitatively, we fit each result to a parabolic function [solid red lines in Fig. 4(b)]. In the fitting procedure, the quadratic term is fixed since carrier doping should not renormalize the band. 
We estimate the conduction band bottom from the fitting curves and plot it against the deposition time in Fig. 4(d) by red dots. 

The potassium deposition introduces not only extra carriers but also Stark effect due to the electric field induced by the ionized potassium adatoms at the surface. Whereas the carrier doping shifts the conduction and valence bands both to higher binding energies, Stark effect shifts these bands in opposite energy directions to each other~\cite{LouiePRB2004,TowePRB2011,KimScience2015,YeomPRL2019,ZhangYPRB2020}. That is why, in our data, the conduction band shifts down and eventually overlaps with the valence band at larger potassium deposition. The shift of the conduction band is confirmed to have a linear behavior with deposition time, as shown by the fitted solid red line in Fig. 4(d).
This implies that the carrier doping and Stark effect both change the band position linearly for energy.
Extrapolation of the conduction band bottom to 0 minutes should remove these linear shifts from the carrier doping and Stark effect, and recover the original band position~\cite{ZhangYPRB2020}. 
However, we found that the bottom of the extrapolated conduction band [blue dashed curve in Fig. 4(d)] does not coincide with that of the original band at 10 K [black solid curve in Fig. 4(d)], which is about 50 meV above $E_\mathrm{F}$. 
Instead, the extrapolated conduction band bottom almost touches $E_\mathrm{F}$. This indicates that there is a third effect that moves down the conduction band at low deposition ($<$ 1 min), except for the carrier doping and Stark effect, which are dominant at high deposition ($>$ 1 min). Since the extrapolated conduction band bottom locates at the energy position the same as the conduction band bottom in the gapless high-temperature phase (350 K), the third effect seems to close the energy gap. Therefore, we conclude that the energy gap is fragile to carrier doping.

\begin{figure}[!htb]
\begin{center}
\includegraphics[width=.45\textwidth]{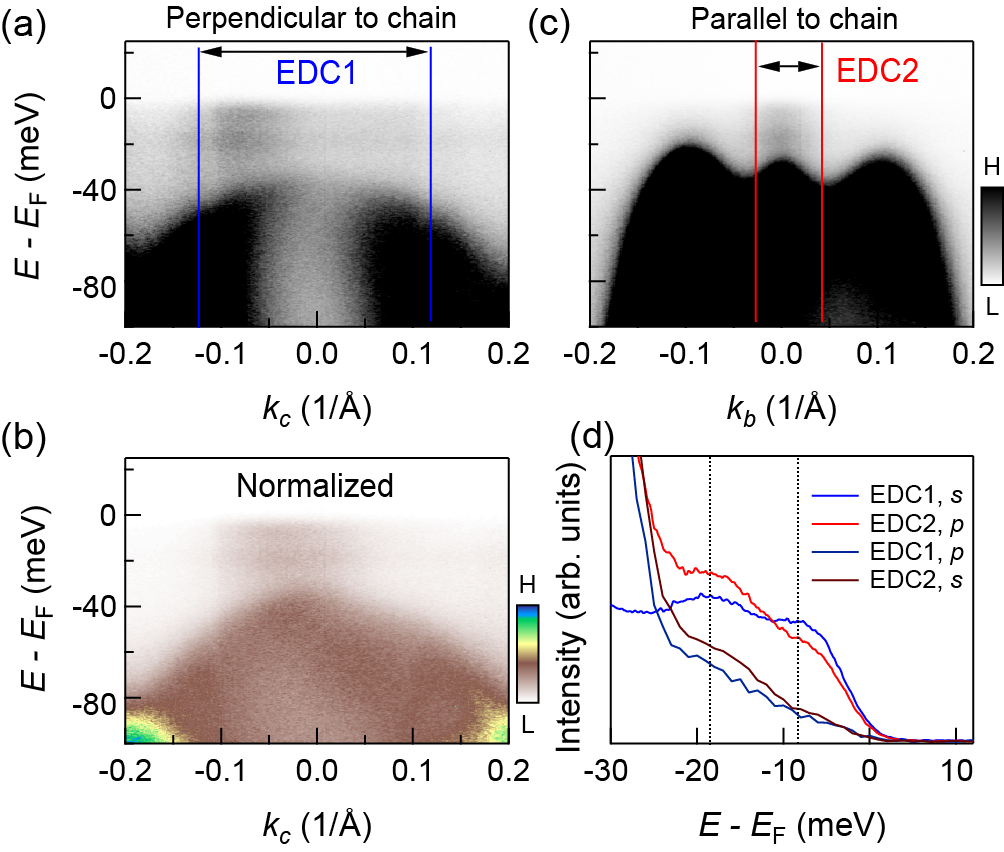}
\end{center}
 \caption{\label{theory} \textbf{Observation of double flat bands inside the energy gap in Ta$_2$Pd$_3$Te$_5$.} (a) Band structure observed perpendicular to the chain direction at $k_b=0$, measured at 10 K with $s$-polarized light. To enhance the visibility of in-gap states, the color scale is adjusted to saturate the intensities of valence bands. (b) The same data as (a) but normalized to the intensities of the valence band to remove the matrix element effect. (c) Band structure observed along the chain direction at $k_c=0$, measured at 10 K with $p$-polarized light.
 (d) EDCs focusing on the in-gap flat bands. EDC 1 and EDC 2 are integrated over momentum regions 
 indicated by the arrows in (a) and (c), respectively. These EDCs are measured with both $s$- and $p$-polarized light. The dashed lines mark peak energies of about -8 and -18 meV.}
\end{figure}

By closely investigating the data measured by high-resolution laser-ARPES, we found double flat bands near $E_\mathrm{F}$ within the energy gap at 10 K. The in-gap flat bands have relatively low spectral intensities, and these are observed only with the light of proper polarization. 
In Figs. 5(a) and 5(c), we exhibit the ARPES images capturing these flat-band signals perpendicular to and parallel to the 1D chains, respectively. The flat bands are clearly visible, although the color contrast needs to be changed till saturating the intensities of the valence band. The spectral weight for the in-gap flat bands modulates along the momentum axis [Fig. 5(a)]. 
This modulation coincides with the momentum variation of the valence band intensity due to the matrix element effect; it is, indeed, removed by normalizing the EDCs to the intensities of the valence band [Fig. 5(b)], which indicates the in-gap flat bands have a close relation with the valence band.
The EDCs of the flat bands are displayed in Fig. 5(d). Here, the spectra are integrated over the momentum regions marked by arrows in Figs. 5(a) and 5(c). The flat bands are located at $\sim$ -8 meV and -18 meV in these data and appear only when using $s$- and $p$-polarized light 
for the data taken perpendicular to and parallel to the chain direction, respectively. 

We note here that these flat bands are relatively weak in intensity but far enough from $E_\mathrm{F}$ as clear from EDCs in Fig. 5(d), so that they are not a residual part of a conduction band cut off by the Fermi function, such as polaronic bands~\cite{MoserPRL2013,BaumbergerNM2016,LouiePRR2020}. The in-gap flat bands generally come from random impurities or defects in materials~\cite{BalatskyPRB1995,SchriefferPRB1997,PanNature2000,BalatskyRMP2006,ZhangPRL2009,ZhangPRX2014}. We notice that the large-scale topographic image of scanning tunneling microscopy shows a reasonable amount of defects in Ta$_2$Pd$_3$Te$_5$~\cite{FengPRB2021}.  
The M-H and susceptibility measurements for our samples (Supplementary Figure 10) show no evidence of magnetic order. By fitting a curve of the paramagnetic susceptibility to the data, we get a magnetic moment per atom to be about 0.03$\mu_B$. This value is small enough to conclude that the impurities/defects do not have a large magnetic moment and should be non-magnetic.  
We also measured crystals synthesized by a different method and detected similar but more intense signals of the in-gap states, where the temperature dependence of the in-gap states also suggests an impurity origin (See Supplementary Figure 7 for more details). In a simple two-band model with local impurity scattering, only one flat band of impurity states exists, either below the conduction band or above the valence band. However, if the conduction and valence bands are hybridized when the gap opens, it is possible to generate double impurity bands inside the gap \cite{IchinomiyaPRB2001,BalatskyPRL2010}. For example, in superconductors, the impurity states appear in a pair inside the superconducting gap due to the particle-hole mixing \cite{BalatskyRMP2006}. Similarly, the double in-gap impurity bands observed in our data suggest that the conduction and valence bands are mixed when the energy gap opens.

The above results on the energy gap can be summarized in the following four points: (i) The gap opens at 350 K from an ideal Dirac semimetal state, and it is associated with a tiny lattice distortion; (ii) The gap opening is absent in a similar material Ta$_2$Ni$_3$Te$_5$. (iii) Potassium deposition destroys the energy gap; (iv) Double in-gap impurity bands are observed, suggesting that the conduction band and valence band are hybridized when the gap opens.  Some interactions involving electrons may cause the opening or enlarging of a band gap (namely,  ``interaction-induced gap/pseudo-gap"). For example, electron scattering can yield a gap-like feature~\cite{KimNature2021}, and electron-phonon coupling can cause a gap size change with temperature in semiconductors~\cite{CohenPRL2010,MonserratPRB2016}.
However, interaction-induced gap/pseudo-gap should not be accompanied with a lattice distortion, should not be fragile to potassium deposition, and should not lead to double impurity states. 
The spontaneous gap opening we observed in Ta$_2$Pd$_3$Te$_5$ is most likely induced by an ordered state (namely, ``order-induced gap''). For example, CDW, spin-density-wave (SDW), superconductivity, and excitonic states, they all generate an electronic energy gap when the order appears. 
Among electronic orders, we exclude the possibility of superconductivity by resistivity measurement.
Density-wave states (CDW/SDW) could not be a candidate either, since the Fermi surface exists only around $k$=0, having no chance to form density waves with a finite wavevector Q.

An excitonic order matches the observed results in a perfect way. First of all, Ta$_2$Pd$_3$Te$_5$ is an almost ideal Dirac semimetal. The carrier density is almost zero and the electron and hole excitations are almost symmetric. Therefore, the screening of the Coulomb interaction is very weak and the electrons and holes can be easily bounded by long-range Coulomb interaction, of which the binding energy is clearly larger than the semimetal gap (zero eV). Thus, an excitonic order could naturally occur.
More importantly, all four observations mentioned above on the energy gap from (i) to (iv) can be explained well by excitonic states:
 (i) The excitonic gap has a similar temperature dependence to that of superconductors or CDW/SDW. When the excitonic order forms, the electron-phonon coupling leads to the tiny lattice distortion. (ii) Ta$_2$Pd$_3$Te$_5$ has an ideal Dirac-type band. Such band structure is very unstable against excitonic order. So it is natural to have excitonic order in Ta$_2$Pd$_3$Te$_5$, and no such order in semiconducting Ta$_2$Ni$_3$Te$_5$. (iii) When extra carriers are doped, the electron-hole balance could be destroyed, and the binding of excitons becomes weak, reducing the excitonic gap in Ta$_2$Pd$_3$Te$_5$. Note that such technique, doping carriers by potassium to destroy the excitonic gap, has been widely used to prove the existence of excitonic states \cite{YeomPRL2019, ZhangYPRB2020, ShenNC2023}. (iv) According to calculations, double flat bands of in-gap impurity states should appear in excitonic insulators with non-magnetic impurities~\cite{ZittartzPR1967,IchinomiyaPRB2001,BalatskyPRL2010}. All these arguments lead us to conclude that an excitonic order is the most likely mechanism for the energy gap in Ta$_2$Pd$_3$Te$_5$. 
In the previous excitonic candidate Ta$_2$NiSe$_5$, there is an obvious structure transition when the gap opens, which leads to the debate whether the gap is originated from structure transition or from excitonic order. In Ta$_2$Pd$_3$Te$_5$, the lattice distortion is very tiny and even undetectable by XRD and specific heat measurements (Supplementary Figure 2). It is unlikely to produce the observed large gap from such a tiny lattice distortion. The spontaneous gap opening should come from the excitonic order, and the tiny lattice distortion is a secondary phenomenon from the excitonic order. 

An ordered state expects a jump in specific heat at the critical temperature. 
However, no jump was distinguished in our specific heat measurement at the gap-opening temperature (350 K), as shown in Supplementary Figure 2. 
We note that a pure electronic transition at high temperature should not be judged from the specific heat jump, as electronic specific heat at high temperature is too small compared to the lattice specific heat. 
For example, the electronic heat capacity constant of iron, a typical metal, is $\gamma$ = 4.98 mJ/(mol K$^2$) \cite{Kittel-book}, and the electronic heat capacity at 350 K is estimated to $C_e=\gamma T$ =1.7 J/mol K. 
On the other hand, a total heat capacity of Ta$_2$Pd$_3$Te$_5$ is obtained as about 250 J/mol K at 350 K from our measurement (Supplementary Figure 2). A simple calculation tells us that the electronic heat capacity is two orders smaller than the lattice heat capacity: 1.7/(250-1.7) $\ll$ 0.01. In particular, Ta$_2$Pd$_3$Te$_5$ is an ideal Dirac semimetal with almost zero carrier density, thus the electronic heat capacity must be even much smaller than that of a typical metal (For example, the electronic heat capacity of semimetallic bismuth is more than 100 times smaller than that of iron \cite{Kittel-book}). 
Therefore, even if there is a jump induced by electronic order, it would be too small to be distinguished at 350 K. Interestingly, this situation is similar in superconductors: the specific heat jump for superconducting transition in some cuprates is not distinguished with $T_c$ around or higher than 100 K~\cite{GauzziPRB2016}.
On the other hand, although there is a lattice distortion accompanied with the excitonic transition resolved by electron diffraction, it is very tiny and likely does not produce an observable specific heat jump either, which we discussed previously. 

The previous excitonic insulator candidates TiSe$_2$ and Ta$_2$NiSe$_5$ coexist with either CDW or structure transition~\cite{CercellierPRL2007,TakagiPRL2009}. The coexistence of different phases makes it difficult to justify excitonic states. 
In contrast, there is no entangling order in Ta$_2$Pd$_3$Te$_5$. The lattice distortion observed in electron diffraction is too tiny and should be a consequence of the excitonic order.
Although further experimental evidence would be desired to reach a definitive conclusion, our results provide that quasi-1D Ta$_2$Pd$_3$Te$_5$ is the first promising candidate of an excitonic insulator in which entangled phases are absent, being highly advantageous for developing the condensed matter physics on excitonic states.  
\newline

\textbf{Acknowledgement} We thank J.X. Li, H.J. Zhang, W. Chen, Z. Zhu, X.X. Wu, Q.H. Wang, S. Bao, J.S. Wen and J.R. Huang for useful discussions and support. This work was supported by the National Natural Science Foundation of China (12274209, U2032204, 11974395, 12188101 and 11504329), the Informatization Plan of Chinese Academy of Sciences (CAS-WX2021SF-0102), the Strategic Priority Research Program of Chinese Academy of Sciences (Grant No. XDB33000000), the China Postdoctoral Science Foundation funded project (Grant No. 2021M703461), the Center for Materials Genome, Zhejiang Provincial Natural Science Foundation of China (Grant No. LZ23A040002), 
the JSPS KAKENHI (grant numbers JP18H01165 and JP19H00651), JST ERATO-FS (grant number JPMJER2105), MEXT Q-LEAP (grant number JPMXS0118068681), and by MEXT under the “Program for Promoting Researches on the Supercomputer Fugaku” (Basic Science for Emergence and Functionality in Quantum Matter Innovative Strongly Correlated Electron Science by Integration of “Fugaku” and Frontier Experiments, JPMXP1020200104) (Project ID: hp220166).

\end{document}